\documentclass[12pt]{iopart} \input epsf
\begin{document}

\title{On nonlinear susceptibility in supercooled liquids}

\author{
Silvio Franz$^1$ and Giorgio Parisi$^2$
}
\address{
$^1$The Abdus Salam ICTP, Strada Costiera 11, P.O. Box 563,
34100 Trieste, Italy\\
$^2$Universit\`a di Roma ``La Sapienza'' P.le A. Moro 2, 00185 Rome,
Italy}

\begin{abstract}
In this paper, we discuss theoretically the behavior of the four point
nonlinear susceptibility and its associated correlation length for
supercooled liquids close to the Mode Coupling instability temperature
$T_c$. We work in the theoretical framework of the glass transition as
described by mean field theory of disordered systems, and the
hypernetted chain approximation. Our results give an interpretation 
framework for recent numerical
findings on heterogeneities in supercooled liquid dynamics.

\end{abstract}


\submitted{{\noindent \it }}

\section{Introduction}

Recently, a lot of attention has been devoted to understanding the
nature of dynamical heterogeneities in supercooled liquids
\cite{donati,hiw,heuer,dgp,bdbg,dgpkp,PARI}.  Many numerical
experiments have found long lived dynamical structures which are
characterized by a typical length and a typical relaxation time which
depend on the values of the external parameters (temperature and
density).  A way to quantify this dynamical heterogeneities is in
terms of the 4-point density function, and its associated non-linear
susceptibility, which show power law behavior as one approaches the
Mode Coupling temperature $T_c$ from above. In this paper we review
the details of the theoretical calculations 
of this function put forward in \cite{comment,prl,andalo} and
discuss some new results.






At the glass transition one observes freezing of density fluctuations.
The function 
\begin{equation}
g_2(x)=\langle (\rho(x+y)-\overline{\rho})(\rho(y)-\overline{\rho})\rangle
\end{equation}
is often regarded as the Edwards-Anderson order parameter signaling
the onset of glassiness. It is therefore quite natural to try to
interpret the dynamical heterogehinities and the correlation length in
terms of fluctuations of the order parameter, and study the 4-point
function
\begin{equation}
g_4(x)=\langle
[(\rho(x+y)-\overline{\rho})(\rho(y)-\overline{\rho})]^2\rangle-\langle
(\rho(x+y)-\overline{\rho})(\rho(y)-\overline{\rho})\rangle^2, 
\end{equation}
and its related non-linear susceptibility $\chi_4=\int dx g_4(x)$.  To
our knowledge the first proposal to study the 4-point function to
identify a growing correlation length in structural glasses was in
\cite{dasg} in the context of a numerical study of a Lennard-Jones
liquid.  There no sign of growing correlation was found, probably
because of the insufficient thermalization. However, more accurate
measurements \cite{dgp,bdbg,prl,shult} show that there is a dynamical
correlation length which grows as $T_c$ is approached.

Here we would like to investigate theoretically the behavior this
function in the context of the picture of the glass transition that
comes out from the study of disordered mean-field models
\cite{kirtir}, and from some approximation scheme of molecular liquids
\cite{pame}.

In mean field disordered systems one finds that decreasing the
temperature from the liquid phase, two different transitions appear: a
dynamical transition at a temperature $T_c$, and a static
(Kauzmann-like) transition at a lower temperature $T_K$.  At the
dynamical transition $T_c$, identified with the Mode Coupling Theory
\cite{MCT} transition temperature, equilibrium density fluctuations
freeze and ergodicity breaks down.  Below that temperature, the
Boltzmann distribution is decomposed in an exponentially large number
of ergodic components $\e^{N\Sigma(T)}$.  $\Sigma(T)$, the logarithm
of the number of these components is the configurational entropy,
which decreases for decreasing temperatures, and the ``static''
transition signals the point where $\Sigma(T_K)=0$.  Dynamically, a
non zero Edwards-Anderson order parameter signals freezing.

As it has been many times remarked, this theory misses the existence
of local activated processes which restore ergodicity below
$T_c$. These can be included phenomenologically to complete the
picture. We will suppose that the ergodic components which the ideal
theory predicts below $T_c$ become in real systems metastable states
(or quasistates), capable to confine the system for some large, but
finite times on given portions of the configuration space.
The inclusion of activated processes, although done by hand, has far
reaching consequences. 

The foundation of the 
notion of quasistates is based on the time scale separation (as it
can be seen in the shape of the structure function), which allows to
consider ``fast'' degrees of freedom quasi-equilibrated, before the
``slow'' degrees of freedom can move. So, this notion 
applies below as well as  above $T_c$, where the 
two step relaxation is predicted even by the ideal theory. This 
point has been recently stressed in \cite{fravi} in a different context. 
Both above and below $T_c$ we can
talk of quasistates in which the system equilibrates almost completely
before relaxing further.  The typical life time of the quasistate will
be of the order of the alpha relaxation time $\tau_\alpha$.

Our basic observation is that within the described theoretical framework, 
the quasistates correspond to highly correlated regions of the
configuration space, typical configurations belonging to the same
quasistate would appear to be highly correlated. 
On the other hand, configurations belonging to 
distant quasistates as the ones which correspond to large time
separation $t>>\tau_{\alpha}$, show typically low correlations. 

We argue then, that the dynamical correlation length and susceptibility
observed in the simulations referred to above, can be estimated by the corresponding 
quantities within a quasistate. On the other hand, the long time limit 
of the same quantities, i.e. the value reached for times 
much larger then the lifetime of the quasistates, 
correspond to maximally distant quasistates. 
This predicts maximal fluctuations and heterogeneity on a time
scale of the order of $\tau_{\alpha}$. 

\section{How to compute quasistate averages: ``recipes for metastable states''}
\label{sec:section1}

In this section, we address the question on how to compute correlation
functions within singles quasistates, reviewing some ``recipes'' 
that were put forward in \cite{pot,hnc}. Let us consider the case
of mean-field spin glass models below $T_c$, where there is true
ergodicity breaking and the quasistates are true ergodic components.
Suppose to be above $T_K$ so that the configurational entropy
$\Sigma(T)>0$. Given any local observable $A(x)$, its Boltzmann average can be
decomposed as 
\begin{equation}
\langle A(x)\rangle_{Boltzmann}=\sum_\alpha w_\alpha \langle
A(x)\rangle_{\alpha},
\label{deco}
\end{equation}
where the index $\alpha$ runs over all the $e^{N\Sigma}$ states, 
the weights of the different states $w_\alpha$ would all be
of the same order $w_\alpha\approx \exp(-N\Sigma(T))$. 
In the following we will be interested to compute space averages
(correlation functions) among local observables,
$\int dx \langle A(x)\rangle \langle B(x+y)\rangle$. 
If by $\langle \cdot \rangle$ we mean Boltzmann average, we can 
expand each of the two averages according to (\ref{deco}) and find 
that 
\begin{equation}\int dx \langle A(x)\rangle \langle B(x+y)\rangle=
\int dx \sum_{\alpha,\beta} w_\alpha w_\beta
\langle A(x)\rangle_\alpha \langle B(x+y)\rangle_\beta,
\label{bb}
\end{equation}
which, due to the fact that 
 the number of ergodic components is exponentially large, 
is dominated by the terms in the double sum with $\alpha\ne
\beta$. 

Our major interest will be to compute instead averages of the kind
$\int dx \sum_\alpha w_\alpha \langle A(x)\rangle_\alpha 
\langle B(x+y)\rangle_\alpha$
i.e. correlation function in the same ergodic component. 
To this purpose we can use a {\it conditional} Boltzmann
prescription \cite{pot,hnc}, where 
one fixes a reference configuration $Y=\{y_1,...,y_N\}$, and 
only the configurations $X=\{x_1,...,x_N\}$ similar 
enough to the reference configuration  are given a
non vanishing weight. 

Let us consider as a notion of similarity 
among two configurations
$X$ and $Y$
the function, that we call overlap,
\begin{equation}
q(X,Y)=\int dx dy 
(\rho_X(x)-\overline{\rho})(\rho_Y(y)-\overline{\rho})w(|x-y|)
\end{equation}
where:
\begin{itemize}
\item
 $\rho_Z(z)$ ($Z=X,Y$) is the microscopic density corresponding
to the configuration $Z$: 
$\rho_Z(z)=\sum_i \delta(z_i-z)$
\item
the function $w(r)$ is a short range sigmoid (or step) function such
that if $r_0$ denotes the typical radius of the particles, $w(r)$ is
close to 1 for $r\le a r_0$ and close to zero otherwise.  The value of
$a=0.3$ gives a notion of overlap not too sensitive to small atomic
displacements.
\end{itemize}
Notice that with this definition $q(X,Y)$ is maximal if $X=Y$, while 
it is equal to zero if $X$ and $Y$ are uncorrelated. 
Notice also that one can write $q(X,Y)=\sum_{i,j}w(|x_i-y_j|)$, a 
form which is manifestly invariant under permutations of the particles. 

Suppose now to fix a reference configuration $Y$, chosen with
Boltzmann probability at temperature $T$, and consider the 
conditional probability 
\begin{equation}
P_q(X|Y)=\frac{e^{-\beta H(X)}\delta(q(X,Y)-q)}{Z_q(Y)}
\label{cond}
\end{equation}
where the constrained partition function is
\begin{equation}
Z_q(Y)={\int d X
e^{-\beta H(X)}\delta(q(X,Y)-q)}. 
\end{equation}
As $Y$ by hypothesis is an equilibrium configuration, it will belong
to some quasistate $\alpha$, so, if we choose $q$ as the typical 
overlap among configurations in this quasistate (with probability one 
almost all configurations have the same overlap), i.e. the
Edwards-Anderson parameter of the state $q_{EA}$, we will be able to 
compute the quasistate averages: 
given two observables $A(X)$ and $B(X)$ we can write:
\begin{equation}
\sum_{\alpha}w_\alpha \langle A\rangle_\alpha \langle B\rangle_\alpha
=
\int d Y \frac{e^{\-\beta H(Y)}}{Z}A(Y) \int dX \frac{e^{-\beta
H(X)}\delta(q(X,Y)-q)}{Z_q(Y)} B(X)
\end{equation}
Notice that, if 
on the other hand in (\ref{cond}) we would choose $q$ as the typical 
overlap among different quasistates, the constraint would be  completely
irrelevant and we could get the Boltzmann average (\ref{bb}).

Notice that the overlap we consider, is a masked integral of 
the density-density correlation function among the two configurations 
$X$ and $Y$. We are interested to study the fluctuation of this
quantity, which is the following integral of the 4-point function:
\begin{eqnarray}
\chi_4=& &\beta(\overline{\langle q \rangle^2}-\overline{\langle q \rangle}^2)
\nonumber
\\ = & &
\int \; dx\; dy\; dz\; dr w(x-y)w(z-r) 
\nonumber
\\
& &\left(
\overline{
\langle 
(\rho_X(x)-\overline{\rho})
(\rho_Y(y)-\overline{\rho})\rangle\langle
(\rho_Y(z)-\overline{\rho})
(\rho_Z(r)-\overline{\rho})\rangle}
\right.
\nonumber
\\ & &
\left. -
\overline{
\langle 
(\rho_X(x)-\overline{\rho})
(\rho_Y(y)-\overline{\rho})
\rangle}
\times
\overline{
\langle
(\rho_X(z)-\overline{\rho})
(\rho_Y(r)-\overline{\rho})\rangle}\right)
\label{mostro}
\end{eqnarray}
where we have denoted with the angular brackets the conditional
average with the distribution (\ref{cond}), and with the bar the
average over the canonical distribution of the reference configuration
$Y$.

Within this formalism, the generating functional of the correlation 
functions is the constrained free-energy 
\begin{equation}
V(q)= -\beta /N \int dY \frac{e^{\-\beta H(Y)}}{Z} \log(Z_q(Y)).
\end{equation}

\begin{figure} 
\hbox to\hsize{\epsfxsize=0.8\hsize\hfil\epsfbox{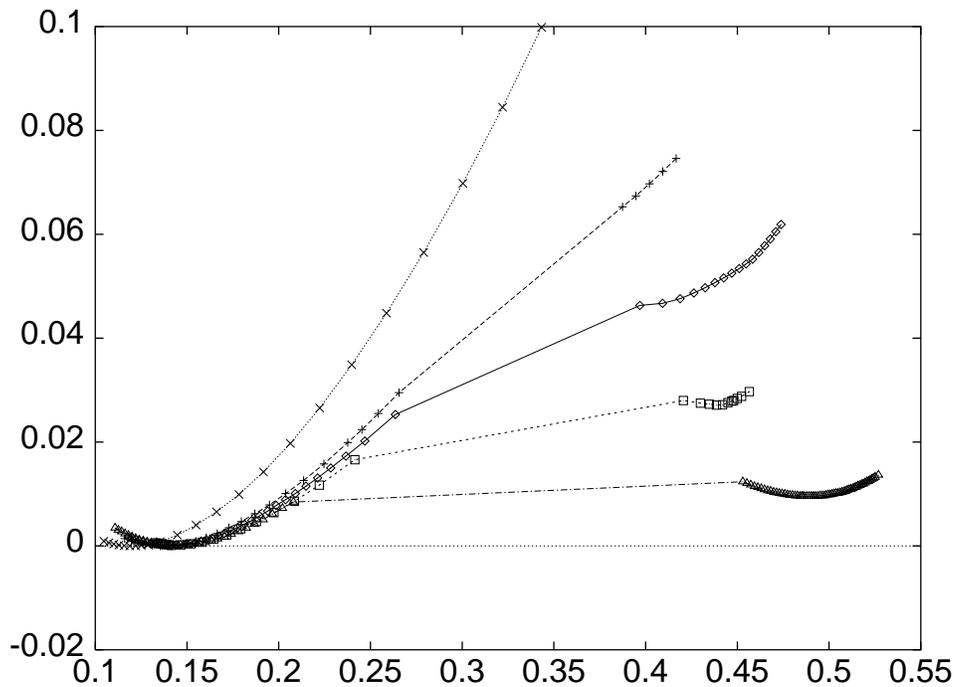}
\hfil}
\caption{The potential for HNC hard spheres at several densities.  We
show the high and low $q$ regions of the potential. The lines joining
them are just guides for the eyes.Lower curves correspond to higher
densities.
At low density the potential is convex. The appearance of a secondary 
minimum signals the breaking of ergodicity, with exponentially many states.}
\label{hnc} 
\end{figure}

This function has been computed in various models having a glass
transition, including mean field disordered models and simple liquids
in the HNC approximation, all giving consistent results
\cite{pot,hnc}. The shape of $V$ as a function of $q$ allows to
distinguish among liquid and glass phase. We show the potential 
 in figure \ref{hnc} for hard spheres in the HNC approximation.


In the liquid phase at high temperature, the potential is convex with
a unique minimum for $q=0$ which corresponds to the typical overlap
among random liquid configurations. Lowering the temperature, the
potential looses the convexity, until, when $T_c$ is reached, it
develops a secondary minimum, at a high value of $q$. The height of
the secondary minimum with respect to the first one is related to the
configurational entropy by: $V_{sec}-V_{pri}=T\Sigma(T)$, which
vanishes at $T_K$.
A remarkable fact that has been often discussed \cite{pot}, is that 
while the properties of the low $q$ minimum reflect the 
properties of the full Boltzmann average, the properties of the 
high $q$ minimum reflect the properties of averages in a single
ergodic component. 

In the shape of the potential the MC transition appears as a spinodal
point and as such it has a divergent susceptibility.  In fact, general
relations in the effective potential theory, 
imply that the susceptibility is given just by the inverse curvature
of the potential in the minimum, i.e. 
 $\chi_4=1/V''(q)|_{secondary \;\; minimum}$. This
quantity, diverges for $T\to T_c$, which, in turn, implies the
divergence of the spatial range of the correlations. Generically, in
all the model studied, the slope of the flex vanishes linearly for
$T\to T_c$, implying $\chi_4(T)\simeq |T-T_c|^{-\gamma}$ with a mean
field exponent $\gamma=1/2$.  We notice that $\chi_4$ computed in the
primary minimum represents the Boltzmann average of the order parameter
fluctuations and is completely regular at $T_c$.

\begin{figure} 
\hbox to\hsize{\epsfxsize=0.8\hsize\hfil\epsfbox{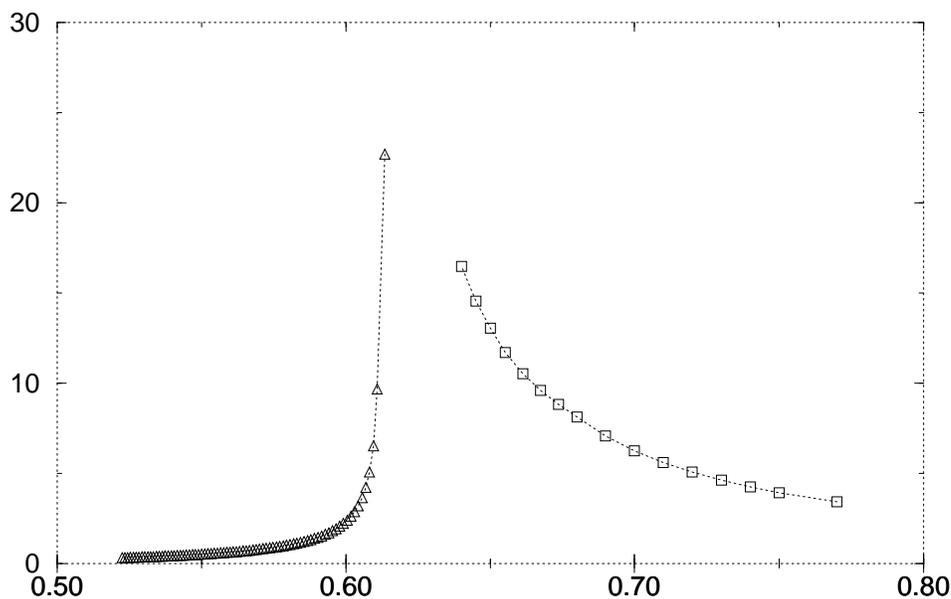}
\hfil}
\caption{The susceptibility of the metastable states for the 
$p$-spin model with $p=3$. Here $T_c=0.612$. The low
temperature data come from the potential theory. The high temperature
data from the dynamical equations.}
\label{figure2} 
\end{figure}

In figure \ref{figure2} we see the susceptibility computed with this
procedure (circles), which shows a divergence at $T_c$.

\section{A dynamical approach}
\label{sec:section2}

The idea of considering a system coupled with a reference
configuration can also be used in dynamics to compute the time
dependent susceptibility.  In this context it is convenient to couple
with the initial configuration $X_0$.  Consider a system at
equilibrium at time zero with respect to the Hamiltonian $H(X)$ which
evolves for positive times with the modified Hamiltonian
\begin{equation}
H_{tot}(X)=H(X)-\epsilon q(X,X_0).
\label{htot}
\end{equation}

For small $\epsilon$, linear response theory at equilibrium implies that 
\begin{equation}
\chi_4(t)=\beta\left( \langle  q(X_t,X_0)^2 \rangle-
 \langle  q(X_t,X_0)q(X_0,X_0) \rangle \right)=\frac {\partial \langle
 q(X_t,X_0) \rangle}{\partial \epsilon}
\end{equation}

The problem of studying the evolution of a system with  
Hamiltonian (\ref{htot}) can in principle be issued within any
dynamical approximation scheme (e.g. MCT). 

However for the time being we have only addressed the problem in the
context of the p-spin model, which, for all the present purposes
should capture the essential features of the function $\chi_4(t)$.
Clearly, lacking completely in the model any spatial structure, in
order to infer from the behavior of $\chi_4$ something about a
correlation length we need to resort to eq. $\ref{mostro}$.

The p-spin model \cite{pspin} describes $N$ interacting variables $S_1,...,S_N$ (spins) 
on the sphere $\sum_i S_i^2=N$, with Hamiltonian
$H=\sum_{i_1<...<i_p}J_{i_1...i_p}S_{i_1}...S_{i_p}$ where the
couplings are random independent Gaussian variables with zero mean and
variance $J^2=p!/(2 N^{p-1})$. The appropriate notion of overlap for
this system
is $q(S,S')=1/N \sum_i S_i S'_i$. For this model it is customary to 
consider Langevin dynamics, which in our case will be performed
with Hamiltonian $H_{tot}(S)=H(S)-\epsilon q (S,S_0)$, where $S_0=S(t=0)$ is 
an equilibrium initial condition. 
\begin{equation}
\frac{d S_i(t)}{dt}=-\frac{\partial H(S(t))}{\partial S_i}+\epsilon S_i(0)
+\eta_i(t)
\end{equation}
where $\eta_i(t)$ is a white noise with amplitude $2T$ and $\mu(t)$
is a Lagrange multiplier which ensures the spherical constraint at all
times.

Using standard manipulations based on the Martin-Siggia-Rose
functional integral, one can write a self consistent equation for a
single spin which, using the notation $f(q)=1/2 q^p$, reads: 

\begin{eqnarray}
\frac{d S(t)}{dt}=& &-\mu(t)S(t)+\int_0^t ds \; f''(C(t,s))R(t,s)S(s)
\nonumber\\
& &+\beta f'(C(t,0))S(0)
+\epsilon S(0)
+\xi_i(t)
\label{leff}
\end{eqnarray}
where 
$\xi(t)$ is a colored Gaussian noise with variance 
\begin{equation}
\langle \xi(t)\xi(s)\rangle =f'(C(t,s))+2T \delta(t-s)
\end{equation}
where $C$ and $R$ are the correlation and response functions of the
system, to be determined self-consistently by $C(t,s)=<S(t)S(s)>$,
$R(t,s)=\langle\frac{\delta S(t)}{\delta \xi(s)}\rangle$.  The
detailed derivation of (\ref{leff}) is rather standard (see
e.g. \cite{horner}) and we do not reproduce it here.

\begin{figure} 
\hbox to\hsize{\epsfxsize=0.8\hsize\hfil\epsfbox{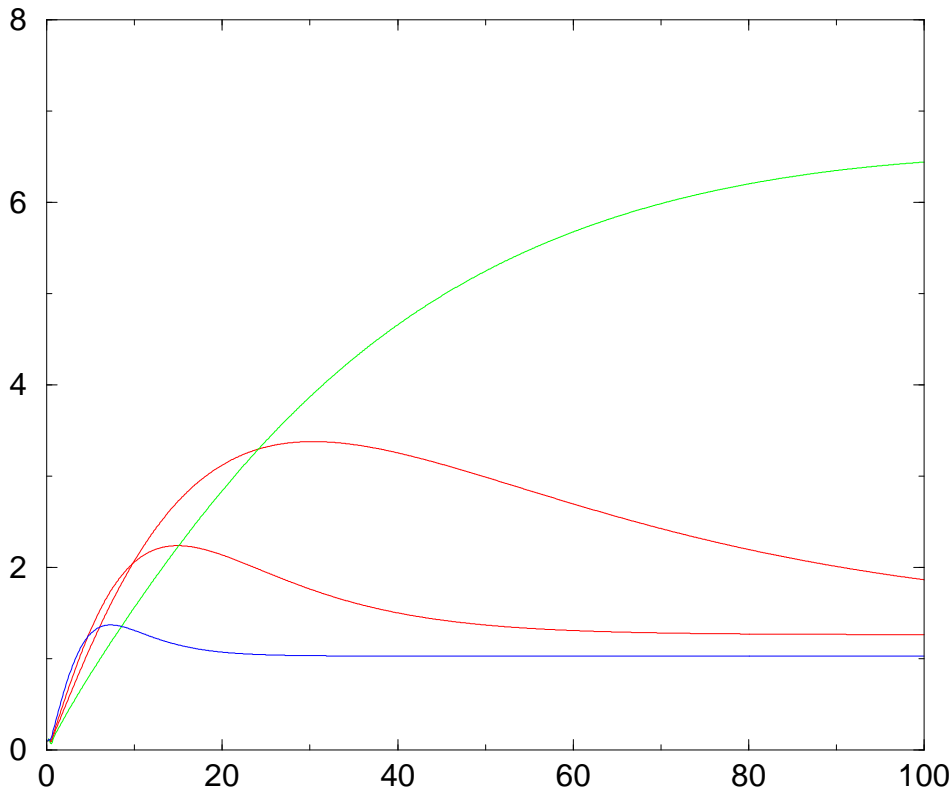}
\hfil}
\caption{The dynamical susceptibility for the p-spin model (p=3)
as a function of time for
several temperatures the lower is the maximum the higher is the temperature. 
From bottom to top $T=1.0,0.9,0.8,0.7$. 
}
\label{figure3} 
\end{figure}

From (\ref{leff}), taking the correlations with $S(s)$ and $\xi(s)$
 one can derive equations for $C$ and
$R$, which read, for $t>s$
\begin{eqnarray}
\frac{\partial C(t,s)}{\partial s}& = &  -\mu(t) C(t,s)+\int_0^t du
f''(C(t,u))R(t,u)C(u,s) \nonumber
\\& &+ 
\int_0^s f'(C(t,u))R(s,u) +
\beta f'(C(t,0)) C(s,0)+\epsilon C(s,0)
\nonumber\\
\frac{\partial R(t,s)}{\partial s}& = & -\mu(t) R(t,s)+\int_s^t du
f''(C(t,u))R(t,u)R(u,s).
\label{cr}
\end{eqnarray}
Together with the equation specifying the time dependence of $\mu(t)$
\begin{eqnarray}
\mu(t)& &=\int_0^t du
f''(C(t,u))R(t,u)C(u,t) \nonumber
\\& &+ 
\int_0^t f'(C(t,u))R(t,u) +
\beta f'(C(t,0)) C(t,0)+\epsilon C(t,0)+T.
\label{mu}
\end{eqnarray}
Equations (\ref{cr},\ref{mu}) form a complete set, that 
can be solved numerically, to derive the value of $\chi_4(t)$ as
\begin{equation}
\chi_4(t)=dC(t,0)/d\epsilon. 
\end{equation}
With a simpl step-by-step integration \cite{frame} we could 
reach times of the order of 1000. This allowed us, in the case $p=3$, to 
compute the function $\chi_4(t)$ down to temperature $T=0.7$, compared
with a critical temperature $T_d=0.612$. 
More sophisticated algorithms
(see the contribution of Latz to these proceedings) will allow
in the next future to approach the critical temperature much more. 
Our results for the function $\chi_4(t)$ are displayed in figure
\ref{figure3} for various temperatures. We see that 
$\chi$ has a maximum which becomes higher and higher as the temperature
is lowered, and is pushed towards larger and larger times. This is 
the behavior that qualitatively is seen in the numerical simulations
\cite{prl,shult}.

As shown in ref. \cite{prl} we define $t^*$ as the time at which
$\chi^*$ is maximum, we find that $\chi^*_4=\chi_4(t^*)$ exhibit a
divergence at $T_c$, as it is presented in figure \ref{figure2}.  Both
quantities behave as powers of $T-T_c$ $t^*\sim (T-T_c)^{-\alpha}$,
$\chi^*_4\sim (T-T_c)^{-\gamma}$.  A best fit gives the values
$\gamma=0.52\pm 0.02$, $\alpha=1.1\pm 0.1$.



\section{Conclusions}
\label{sec:conclusions}

In this paper we have reviewed the analysis of the 
non linear susceptibility in supercooled liquids and glasses 
that comes from the mean field theory of disordered systems, and 
liquid models in the HNC approximation. 

The theory predicts that while the long time, equilibrium
susceptibility remains finite and is regular at all temperatures, the
finite time susceptibility displays a maximum as a function of time
which becomes higher and higher and displaced to larger and larger
times for temperatures close to $T_c$. This behavior is a consequence
of the critical character of the Mode Coupling like dynamical
transition predicted by the ideal theory described in this paper.  I
real systems one can expect a similar behavion but with a round off of
the divergence.

\section{Acknowledgments}
\label{acknowledgements}
We thank C. Donati and S. Glotzer for many discussions on the
topics of this paper.


\section{References}
\begin{thebibliography}{999}

\bibitem{donati} W.Kob, C.Donati, S.J.Plimpton, P.H. Poole and
S.C. Glotzer, Phys. Rev. Lett. {\bf 79} (1997) 2827.  C. Donati,
J.F. Douglas, W. Kob, S.J. Plimpton, P.H. Poole and S.C. Glotzer, 
Phys. Rev. Lett. {\bf 80} (1998) 2338.

\bibitem{hiw} Y. Hiwatari and T. Muranaka, {\it J. Non-Cryst. Sol.}
{\bf 235-237}, 19 (1998); D. Perera and P. Harrowell, {\it ibid},
314.;
 A. Onuki and Y. Yamamoto, {\it ibid}, 34. 

\bibitem{heuer} B. Doliwa and A. Heuer, Phys. Rev. Lett. {\bf
80}, 4915 (1998).

\bibitem{dgp}C. Donati, S. C. Glotzer and P. H. Poole,
Phys. Rev. Lett. in press.

\bibitem{bdbg} C. Benneman, C. Donati, J. Baschnagel and S.C. Glotzer, 
Nature, {\bf 399}, 246 (1999).  See also p.~207. 

\bibitem{dgpkp} C. Donati, S.C. Glotzer, P.H. Poole, W. Kob, and S.J. Plimpton,
Phys. Rev. E {\bf 60} 3102 (1999).

\bibitem{PARI} G. Parisi, J. Phys. Chem, {\bf 103}(20), 4128 (1999).

\bibitem{comment} S. Franz and G. Parisi, unpublished comment
cond-mat/9804084.

\bibitem{prl} C.Donati, S.Franz,  G.Parisi, S.C. Glotzer, {\it Theory of
Non-linear Susceptibility and Correlation Length in Glasses and
Liquids} Preprint cond-mat/9905433

\bibitem{andalo} S.Franz, C.Donati, G.Parisi and S.C. Glotzer,
Phyl. Mag. B {\bf 79} 1827 (1999). 

\bibitem{dasg}C. Dasgupta, A.V. Indrani, S. Ramaswami,  and
M.K. Phani, Europhys. Lett. {\bf 15} (1991) 307 [Addendum:
Europhys. Lett. {\bf 15} (1991) 467].  

\bibitem{shult}S.C. Glotzer, V.N. Novikov, T.B. Schroeder 
preprint cond-mat/9909113

\bibitem{kirtir} T.R. Kirkpatrick and P.G. Wolynes, Phys. Rev. {\bf A
35}, 3072 (1987); T.R. Kirkpatrick and D. Thirumalai, Phys.  Rev. {\bf
B 36}, 5388 (1987); T.R. Kirkpatrick and P. G. Wolynes,
Phys. Rev. {\bf B 36}, 8552 (1987). 

\bibitem{pame} M. Mezard and G. Parisi, J. Phys. A {\bf 29} 65155
(1996).

\bibitem{MCT} W. G\" otze, {\it Aspects of the Structural Glass
Transition} in {\it Liquids, Freezing and the Glass Transition} 
J.P. Hansen, D. Levesque, J. Zinn-Justin eds. North Holland 1990.


\bibitem{fravi} S. Franz and M. A. Virasoro {\it Quasi-equilibrium
interpretation of aging dynamics} preprint cond-mat/9907438.

\bibitem{pot} S. Franz and G. Parisi, J.Physique I {\bf 5} (1995)
1401; Phys. Rev. Lett. {\bf 79} (1997) 2486; Physica A {\bf 261}, 317
(1998).

\bibitem{hnc} M. Cardenas, S. Franz, G. Parisi, J.Phys. A:
Math. Gen. 31  L163 (1998); J. Chem. Phys. {\bf 110},  1726 (1999).

\bibitem{pspin} A review of the $p$-spin model can be found in
A. Barrat cond-mat/9701031 (unpublished).

\bibitem{horner} A. Crisanti, H. Horner and H.J. Sommers, Z. Phys. B {\bf 92},
257 (1993).

\bibitem{frame} S. Franz and M. Mezard, Physica A {\bf 210} (1994)
48. 














 














\end{thebibliography}
\end{document}